\documentstyle [12pt]{article}
\begin{document}
\def\xc12pg{$^{12}C(p,\gamma)^{13}N$}
\def\c12ag{$^{12}C(\alpha,\gamma)^{16}O$}
\def\xn13pg{$^{13}N(p,\gamma)^{14}O$}
\def\be7pg{$^7Be(p,\gamma)^8B$}
\def\o14{$^{14}O$}
\def\b8{$^8B$}
\def\n16{$^{16}N$}
\thispagestyle{empty}
\hspace{4in} {\bf UConn-40870-0008} \\
\begin{center}
{\Large{\bf NUCLEAR ASTROPHYSICS WITH SECONDARY (RADIOACTIVE) BEAMS}}
\footnote{Work Supported by USDOE Grant
No. DE-FG02-94ER40870}
\   \\
\   \\
\   \\
{{\bf Moshe Gai}}\footnote{Invited Talk presented at the International Workshop
on Nuclear Physics, Oaxtepec, Mexico, 4-7 January 1995}
\\
{\normalsize{Dept. of Physics, U46, University of Connecticut,
2152 Hillside Rd., \\ Storrs, CT 06269-3046, USA; \ \ GAI@UConnVM.UConn.Edu}}
\   \\
\   \\
\   \\
{\normalsize{\bf ABSTRACT}}
\end{center}
Some problems in nuclear astrophysics are discussed with emphasize
on the ones central to the field which were not solved over the last two
decades, including Helium Burning in massive stars (the
$^{12}C(\alpha ,\gamma)^{16}O$ reaction)
and the $^8B$ Solar Neutrino Flux Problem
(the $^{7}Be(p,\gamma)^{8}B$ reaction).
We demonstrate that a great deal of progress was achieved by
measuring the time reverse process(es):
the beta-delayed alpha-particle emission of $^{16}N$ and the Coulomb
dissociation of $^8B$, using
{\bf Radioactive Beams} (of $^{16}N$ and $^8B$).  In this way
an amplification of the sought for cross section was achieved,
allowing a measurement of the small cross section(s) of relevance for
stellar (solar) processes.

\begin{center}
\bf RESUMEN.
\end{center}
Se discuten algunos problemas de astrof\'isica nuclear con \'enfasis en
aquellos que son centrales de este campo que no han sido resueltos en las
\'ultimas dos d\'ecadas, incluyendose el consumo de Helio en las estrellas
masivas (la reacci\'on $^{12}C(\alpha ,\gamma)^{16}O$ ) y el problema del flujo
de Neutrinos Solares de $^8B$ (la reaccion $^{7}Be(p,\gamma)^{8}B$).
Demostramos que se han hecho grandes progresos mediante la medida de los
procesos invertidos en el tiempo:
la emisi\'on de part\'icula alfa retardada por beta de $^{16}N$ y la
disociaci\'on coulombiana de
$^8B$, usando {\bf Haces Radioactivos} (de $^{16}N$ y $^8B$). En esta forma
se logro una ampliaci\'on de la b\'usqueda de la secci\'on, permitiendo
la medida de seccion(es) peque\~na(s) de relevancia para procesos estelares
(solares).

\newpage

\section{Introduction}

Nuclear Astrophysics, the study of Nuclear inputs to astrophysics theories,
is a mature science \cite{Fo84} that has developed to the point where
we can now use stars to probe fundamental physics questions.
However, as some critical problems are left unsolved, it presents
a formidable challenge to Nuclear Science.  In
this paper we address two such problems involving key reaction rates:
the \c12ag and \be7pg, of importance for understanding helium burning
and \b8 solar neutrino flux, respectively. We demonstrate that a great deal
of progress was achieved by using {\bf Radioactive Beams} to study
these reactions in their time reverse process.  These new techniques
allow us to add useful data and constraints on the studied nuclear
processes.

\section{Nucleosynthesis in Massive Stars}

Stars commence their energy generating life cycle by burning hydrogen to
form helium.
As stars consume their hydrogen fuel in the core, now composed mainly of
helium, it contracts, raising its temperature and density.  For example, in
25 solar masses stars the hydrogen burning lasts for 7 Million years.  At
temperatures of the order of 200 MK \cite{We80,Ro88}, the
burning of helium sets in.
The first reaction to occur in helium burning is the $\alpha + \alpha
\rightarrow \ ^{8}Be$,
and due to the short
lifetime of $^{8}Be$  this reaction yield a small concentration of $^{8}Be$
nuclei in
the star.  However, this reaction is very crucial as a stepping stone for
the next reaction that is loosely described as the three alpha-capture
process: $^{8}Be(\alpha,\gamma)^{12}C$.
The formation of small concentration of $^{8}Be$,   allows for a larger phase
space
for the triple alpha-capture reaction to occur.  This reaction was
originally proposed by sir Fred Hoyle in the 50's, as a solution for bridging
the gap over
the mass 5 and 8, where no stable elements exist, and therefore the
production of heavier elements. The triple alpha capture reaction
is governed by the excited  $0^{+}$  state in $^{12}C$   at 7.654 MeV.
This state was predicted by Fred Hoyle, prior to its discovery (by Fred
Hoyle himself) at the Kellog radiation lab \cite{Fo84}.  One loosely referrs
to this  $0^{+}$  state as the reason for our existence, since without it
the universe will have a lot less carbon and indeed a lot less heavy
elements, needed for life.  Extensive studies of properties of this state
by nuclear spectroscopists, allow us to determine the triple alpha-capture
rate with accuracy better then 10\%.

At the same temperature range (200 MK), the produced $^{12}C$
nuclei can undergo subsequent alpha-particle capture to form $^{16}O$  via
the $^{12}C(\alpha,\gamma)^{16}O$  reaction.
Unlike the triple alpha-capture reaction this reaction occurs in the
continuum.  This reaction is governed by the quantum
mechanical tail of the bound 2$^+$ at 6.92 MeV, and the
interference of the tail of the bound $1^{-}$  state at 7.12 MeV and
the tail of the quasi-bound $1^{-}$  state at 9.63 MeV in $^{16}O$.
These effects eluded measurements of
the cross section (and the S-factor $= E \sigma exp(2 \pi \eta$)  and
$\eta = e^{2}Z_1 Z_2/ \hbar v$)
of the $^{12}C(\alpha,\gamma)^{16}O$ reaction
for the last two decades, in spite of
repeated attempts \cite{Dy74,Ket82,Red87,Kr88,Ou92},
and was only recently measured using the time reverse
process of the disintegration of $^{16}O$,  populated in the beta-decay of
$^{16}N$
\cite{Zh93,Zh93a,Zh93b,Zh93c,Bu93,Az94}.  Helium
burning lasts for approximately
500,000 years in a 25 solar masses star \cite{We80,Ro88}, and occurs at
temperatures of approximately 200 $MK$.

Stars of masses smaller then approximately 8 solar masses will complete
their energy generating life cycle at the helium burning cycle.  They will
be composed mainly of carbon and oxygen and contract to a dwarf, lying
forever on the left bottom corner of the H-R diagram.  More massive stars
at the end of helium burning, commence carbon burning at a temperature of
approximately 600-900 MK.  Carbon burning lasts for 600 years in 25
solar
masses stars \cite{We80,Ro88}.  The main reaction in carbon burning is
the
$^{12}C(^{12}C,\alpha)^{20}Ne$  reaction, but elements such as
$^{23}Na$  and some $^{24}Mg$  are also
produced.  At temperatures of approximately 1.5 BK  (approximately 150
keV) the tail of the Boltzmann distribution allows for the photo-disintegration
of
$^{20}Ne$  with an alpha-particle threshold as low as 4.73
MeV.  This reaction $^{20}Ne(\gamma,\alpha)^{16}O$  serves as a source of
alpha-particles which
are then captured by $^{20}Ne$  to form $^{24}Mg$  and $^{28}Si$.  The neon
burning cycle
lasts for 1 year in a 25 solar masses stars.  These alpha-particles could
also react with $^{22}Ne$, as suggested by Icko Iben [Ib75], to yield neutron
flux via the $^{22}Ne(\alpha,n)^{25}Mg$  reaction and give rise to the slow
capture of
neutrons and the production of the heavy elements via the (weak) s-process.  At
this point the core is rich with oxygen, and it contracts further and the
burning of oxygen commence at a temperature of 2 BK, mainly via the
reaction $^{16}O(^{16}O,\alpha)^{28}Si$, with the additional production of the
elements
sulfur and potassium.  The oxygen burning period lasts for approximately 6
months in a 25 solar masses star.  At temperatures of approximately 3
BK a
very brief (one day or so) cycle of the burning of silicon commence.  In
this burning period elements in the iron group are produced.  These
elements can not be further burned as they are the most bound (with binding
energy per nucleon of the order of 8 MeV), and they represent the ashes of
the stellar fire.  The star now resemble the onion like structure shown in
Fig. 1.

As the inactive iron core aggregates mass it reaches the Chandrasekar limit
(close to 1.4 solar mass) and it collapses under its own gravitational
pressure, leading to the most spectacular event of a supernova.  During a
supernova the electrons are energetic enough to undergo electron capture
by the nuclei and all protons are transposed to neutrons, releasing the
gravitational binding energy (of the order of 3/5GM$^2/R \approx 3*10^{53}$
ergs)
mostly in the form of neutrino's of approximately 10 MeV (and temperature
of approximately 100 BK).  As the core is now composed of compressed
nuclear matter (several times denser than nuclei), it is black to
neutrino's (i.e. absorbs the neutrino's) and a neutrino bubble is formed
for approximately 10 sec, creating an outward push of the remnants of the
star.  This outward push is believed (by some) to create the explosion of a
type II supernova.  During this explosion many processes occur, including
the rapid neutron capture (r process) that forms the heavier elements of
total mass of approximately $M \approx \mu M_\odot$.

The supernova explosion ejects into the inter-stellar medium its ashes from
which at a later time "solar systems" are formed.  Indeed the death of one
star gives the birth of another.  At the center of the explosion we find a
remnant neutron star or a black hole.

It is clear from Fig. 1, that if in the process of helium burning mostly
oxygen is formed, the star will be able to take a shorter route to the
supernova explosion.  In fact if the carbon to oxygen ratio at the end of
helium burning in a 25 solar masses star, is smaller then approximately $15\%$
\cite{We93}, the star will skip the carbon and neon burning and directly
proceed
to the oxygen burning.
For a cross section of the $^{12}C(\alpha,\gamma)^{16}O$  reaction that is
twice
the
accepted value (but not 1.7 the accepted value), a 25 solar masses star
will not produce $^{20}Ne$, since carbon burning is essentially turned off.
This indeed will change the thermodynamics and structure of the core of the
progenitor star and in fact such an oxygen rich star is more likely to
collapse into a black hole \cite{We93} while carbon rich progenitor stars is
more likely to leave behind a neutron star.  Hence one needs to know the
carbon to oxygen ratio at the end of helium burning (with an accuracy of
the order of $15\%$) in order to understand the fate of a dying star and the
heavy
elements it produces.

Since the triple alpha-particle capture reaction: $^{8}Be(\alpha,\gamma)^{12}C$
is very well
understood, see above, one must measure the cross section of the
$^{12}C(\alpha,\gamma)^{16}O$  reaction with high accuracy of the order of
$15\%$  or better.
Unfortunately, this task was not possible
over the last two decades using conventional techniques and was only
recently tackled using radioactive
beams \cite{Zh93,Zh93a,Zh93b,Zh93c,Bu93,Az94}.
This cross section is needed to be measured at the most effective energy for
helium burning (the Gamow window) of 300 keV.  At this energy one may
estimate \cite{Fo84} the cross section to be $10^{-8}$  nbarn, clearly non
measurable
in laboratory experiments.  In fact the cross section could be measured
down to approximately 1.0 MeV and one needs to extrapolate further down to 300
keV.
As we discuss below the extrapolation to 300 keV,
which in most other cases in nuclear astrophysics could be performed
with certain reliability, is made difficult by a few effects.

The cross section at astrophysical energies has contribution from the p and
d waves and is dominated by tails of the two bound states of $^{16}O$, the
$1^-$
 at
7.12 MeV (p-wave) and the $2^+$  at 6.92 MeV (d-wave).  The p-wave
contribution arises from a detailed interference of the tail of the bound
$1^-$  state at 7.12 MeV and the broad $1^-$  state at 6.93 MeV.  The
contribution of the bound $1^-$  state arises from its virtual alpha-particle
width, that could not be reliably measured or calculated.  Furthermore, the
tails of the quasi-bound and bound $1^-$  states interfere in the continuum and
the mixing phase can not be determined from existing data
measured only at higher energies and therefore it does not show sensitivity
to the above questions.  Due to these reasons the cross section of the
$^{12}C(\alpha,\gamma)^{16}O$  reaction could not be measured in a reliable way
at 300 keV,
and the p-wave S-factor at 300 keV, for example, was estimated to be
between 0-500 keV-barn with a compiled value of $S_p(300) = 60 +60 -30$  keV-b
\cite{Ca88,Ba92} and
$S_d(300) = 40 +40 -20$  keV-b.  This large uncertainty is contrasted by the
astrophysical
need to know the S-factor with $15\%$  accuracy.

The beta-delayed alpha-emission of $^{16}N$  allows the study the
$^{12}C(\alpha,\gamma)^{16}O$  reaction in its time reverse fashion, the
disintegration of
$^{16}O$  to $\alpha + ^{12}C$, and it provides a high sensitivity for
measuring
low energy
alpha-particles and the reduced (virtual) alpha-particle width of the bound
$1^-$  state in $^{16}O$  at 7.12 MeV.  As shown in Fig. 2, low energy
alpha-particle emitted from $^{16}N$  correspond to high energy beta's and thus
to a larger phase space and enhancement (proportional to the total energy to
approximately the fifth power).  In addition the apparent larger matrix
element of the beta decay to the bound $1^-$  state provides further
sensitivity to that state.

It is clear from Fig. 2 that the beta-decay in this case provides {\bf "NARROW
BAND WIDTH HI FI AMPLIFIER"}, where the high fidelity is given by our
understanding of the predicted shape of the beta-decay's.  However, in this
case one needs to measure the beta decay with a sensitivity for
a branching ratio of the order of $10^{-9}$  or better.  In this case we have
replaced an impossible experiment (the measurement of the
$^{12}C(\alpha,\gamma)^{16}O$
reaction at low
energies) with a very hard one (the beta-delayed alpha-particle emission of
$^{16}N$).

Prediction of the shape of the spectra of delayed alpha-particles from $^{16}N$
were first published by Baye and Descouvemont \cite{Bay88}, with an
anomalous interference structure around 1.1 MeV, at
a branching ratio at the level of $10^{-9}$.  The reduced alpha-particle width
of the bound $1^-$  state can be directly "read off" the spectra if measured at
such low energies.  We emphasize that these predictions were
published approximately five years prior to the observation of the anomaly
interference structure around 1.1 MeV \cite{Bu93,Zh93}.  The previously
measured beta-delayed alpha-particle emission of $^{16}N$  \cite{Ne74} was
analyzed
using R-matrix theory by Barker \cite{Ba71} and lately by Ji, Filippone,
Humblet
and Koonin \cite{Ji90}.  They conclude that the data
measured at higher energies is dominated by the quasi bound $1^-$ state in
$^{16}O$  at
9.63 MeV and shows little sensitivity to the interference
with the bound $1^-$  state.  The data measured at low energies is predicted to
have large sensitivity to the anomalous interference with the bound $1^-$
state.  Similar prediction were also given by a K-matrix
analysis of Humblet, Filippone, and Koonin \cite{Hu91} of the
same early data on $^{16}N$  \cite{Ne74}.  We again note that both the R-matrix
paper and K-matrix paper were published three and two years, respectively,
prior to publication of the spectra from $^{16}N$  \cite{Bu93,Zh93}.

As shown in Fig. 2, the beta decay can only measure the p-wave S-factor of
the $^{12}C(\alpha,\gamma)^{16}O$  reaction, and it also includes (small)
contribution from an
f-wave.  The contribution of the f-wave has to be determined empirically
and appears to be very small and leads to some (at most $15\%$) uncertainty in
the quoted S-factor \cite{Zh93,Zh93a,Zh93c}.  The extraction of
the total
S-factor of the $^{12}C(\alpha,\gamma)^{16}O$  reaction could then be performed
from the
knowledge of the E2/E1 ratio which is better known then the individual
quantities.

An experimental program to study the beta-delayed alpha-particle emission
of $^{16}N$  (and other nuclei) was commenced at Yale University in early 1989.
A similar program commenced at about the same time at the {\bf TISOL
radioactive beam facility at TRIUMF}.
After some four years of studies and background reduction, the first
observation of the interference anomaly was carried out in November of
1991, and presented by Zhiping Zhao in a seminar at Caltech in January
1992.  This preliminary report of the anomaly around 1.1 MeV with small
statistics (approximately 25 counts in the anomaly), has activated the
TRIUMF collaboration including Charlie Barnes of Caltech, who
redesigned their unsuccessful search using a superconducting solenoid to
remove beta's, and indeed in March of 1992 they also observed 9 counts in
the anomalous structure as reported by Charlie Barnes in the meeting
of the AAAS in Ohio, June, 1992.  The two collaboration have then carried out
the
required checks and balances and analyses of the data and in November 1992,
both collaborations submitted their papers for publication in the Phys.
Rev. Lett. within ten days \cite{Bu93,Zh93}.  While the two experiments
are very different in their
production of $^{16}N$  and detection method, and hence
acquire different systematical error(s), they appear to be in agreement
and quote similar S-factors, of
similar accuracy and with good agreement.  The R-Matrix analysis of the two
experiments \cite{Zh93,Po93,Az94} yield: \\
\   \\
\underline{Yale Result:}  S$_{E1}$(300) = 95 $\pm$ 6 (stat)
   $\pm$  28 (syst) keV-b \\
\   \\
\underline{TRIUMF II Result:}         = 79 $\pm$ 21  keV-b \\
\   \\
In the Yale experiment \cite{Zh93,Zh93a}, recoiling $^{16}N$  nuclei produced
with a 9 MeV deuterium beams with the $^{15}N(d,p)^{16}N$  reaction,
were collected in a $Ti^{15}N$  foil (with Au backing), tilted at $7^{\circ}$
with respect to the beam.  The foil is then rotated using a stepping motor
to a counting area where the time of flight between alpha-particles,
measured with an array of 9 Si surface barrier detectors, and beta-particles,
measured with an array of 12 plastic scintillator, is recorded.  The
experiment is described in details \cite{Zh93,Zh93a,Zh93c} and
it arrived at a
sensitivity for the branching ratio of the beta-decay in the range of $10^{-9}$
(to $10^{-10}$), see Fig. 3, where we show one of the cleanest spectrum.  Other
spectra show background at the level of branching ratio of $10^{-9}$.

The data measured at Yale University arise from alpha-particle that
traverse the production foil and thus need to be corrected for such
effects.  Hence, we have measured the spectra of beta-delayed
alpha-article emission of $^{16}N$  in the absence of foil.  This was achieved
by implanting radioactive $^{16}N$  beams from the
{\bf MSU A1200 radioactive beam
facility} \cite{Zh93b}, into a surface barrier detector and by studying the
alpha-ecay in the detector.  In this experiment the absolute branching ratio
for
the beta decay to the quasi bound $1^-$  state was also measured.

The spectrum after correction for the foil thickness and response function
was fitted with the R-matrix formalism developed by Ji, Filippone, Humblet,
and Koonin \cite{Ji90} as shown in Fig. 4a and 4b. Indeed this spectrum
appears to be in agreement with the one measured at TRIUMF \cite{Bu93} and
more recently at Seattle \cite{Zh95}, as shown in Fig. 5.

An attempt to extract the total (E1 + E2) S-factor for the
$^{12}C(\alpha,\gamma)^{16}O$
reaction from the abundance of the elements was carried out by Weaver and
Woosley \cite{We93}, by comparing the calculated abundance of the elements to
the solar abundance.  In this calculation the $^{12}C(\alpha,\gamma)^{16}O$
cross section is
varied between 0.5 to 3 times the value tabulated in CF88 \cite{Ca88} and
listed
in \cite{Ba92}: $S_{E1} = 60 +60 -30$  keV-b and $S_{E2} = 40 +40 -20$  keV-b.
As shown in
Fig. 6, they can reproduce the observed solar abundances only for a cross
section which is 1.7 $\pm$ 0.5 times CF88.  If we assume the ratio E1/E2 = 1.5,
as suggested in CF88 \cite{Ca88,Ba92}, we derive: \\
\   \\
\underline{Weaver and Woosley's Result:}  S$_{E1}$(300) =
   102 $\pm$ 30 keV-b \\
\   \\
in excellent agreement with laboratory measurements, see above.

\section{The Coulomb dissociation of $^8B$ and the $^8B$  solar neutrino flux}

The Coulomb Dissociation \cite{Bau86} Primakoff \cite{Pr51} process, is the
time
reverse process of the radiative capture.  In this case instead of studying
for example the fusion of a proton plus a nucleus (A-1), one studies the
disintegration of the final nucleus (A) in the Coulomb field, to a proton
plus the (A-1) nucleus.  The reaction is made possible by the absorption of
a virtual photon from the field of a high Z nucleus such as $^{208}Pb$.  In
this
case since $\pi/k^2$  for a photon is approximately 1000 times larger than that
of a particle beam, the small cross section is enhanced.  The large virtual
photon flux (typically 100-1000 photons per collision) also gives rise to
enhancement of the cross section.  Our understanding of QED and the virtual
photon flux allow us (as in the case of electron scattering) to deduce the
inverse nuclear process.  In this case we again construct a {\bf "NARROW BAND
WIDTH HI FI AMPLIFIER"} to measure the exceedingly small nuclear cross
section of interest for nuclear astrophysics.  However in Coulomb
dissociation since $\alpha Z$  approaches unity (unlike the case in electron
scattering), higher order Coulomb effects (Coulomb post acceleration) may
be non-negligible and they need to be understood
\cite{Ba93,Ber94,Typ94}.  The success of
the experiment is in fact hinging on understanding such effects and
designing the kinematical conditions so as to minimize such effects.

Hence the Coulomb dissociation process has to be measured with great care
with kinematical conditions carefully adjusted so as to minimize nuclear
interactions (i.e. distance of closest approach considerably larger then 20
fm, or very small forward angles scattering), and measurements must be
carried out at high enough energies (many tens of MeV/u) so as to maximize
the virtual photon flux.  Indeed when such conditions were not
carefully selected \cite{He91,Gaz92} the measured cross sections were
found to be
dominated by nuclear effects, which can not be reliably calculated to allow
the extraction of the inverse radiative capture cross section.

Good agreement between measured cross section of radiative capture through
a nuclear state, or in the continuum, was achieved for the Coulomb
dissociation of $^6Li$  and the $^4He(d,\gamma)^6Li$  capture reaction
\cite{Ki91}, and the
Coulomb dissociation of $^{14}O$  and the $^{13}N(p,\gamma)^{14}O$  capture
reaction \cite{Mo91,De91,Kie93}.
In addition we note that a test experiment on the Coulomb
dissociation of $^{13}N$  \cite{Mo91} was also found to be in agreement with
the
$^{12}C(p,\gamma)^{13}N$  capture reaction.

The Coulomb dissociation of $^8B$ may provide a good opportunity for resolving
the issue of the absolute value of the cross section of the $^7Be(p,\gamma)^8B$
reaction.  The Coulomb dissociation yield
arise from the convolution of the inverse nuclear cross section times the
virtual photon flux.  While the first one is decreasing as one approaches
low energies, the second one is increasing (due to the small threshold of
137 keV).  Hence as can be seen in Fig. 7 over the energy region of 400 to
800 keV the predicted measured yield is roughly constant.  This is in
contrast to the case of the nuclear cross section that is dropping very
fast at low energies, see Fig. 7.  Hence measurements at these energies
could be used to evaluate the absolute value of the cross section.

The Coulomb Dissociation process measured at lower energies
is insensitive to the M1 component of the
cross section, since the M1 virtual photon flux is smaller by a factor
(smaller than) $\beta ^2$.  In this case the $1^+$ state in $^8B$  at E$_{cm}$
=
632 keV,
is not expected to be observed.  The M1
contribution is predicted to be approximately $10\%$  of the
$^7Be(p,\gamma)^8B$
cross section \cite{Ki87} measured just below 1 MeV, which yield to a $10\%$
correction of the S$_{17}$-factor extracted
from the radiative capture work \cite{Ro73}, but not
the S$_{17}$ factor extracted from the Coulomb dissociation.

An experiment to study the Coulomb dissociation of $^8B$  was performed during
March-April, 1992, at the {\bf RIKEN-RIPS radioactive beam facility}.
The experiment is a Rikkyo-RIKEN-Yale-Tokyo-LLN
collaboration \cite{Mo93}.  The radioactive
beams extracted from the RIPS separator are shown in Fig. 8.
Indeed the results of the experiment allow us to measure the radiative
capture $^7Be(p,\gamma)^8B$  cross section and preliminary results
are consistent with the absolute value of the cross section
measured by Filippone et al. \cite{Fi83} and by Vaughn et al. \cite{Va70},
as shown in Fig. 9.

\subsection{Is There Evidence for an E2 Component?}

The much publicized \cite{Ba94} paper of Langanke and Shoppa (LS) \cite{Lang1}
claims that the data analysis performed by the RIKEN collaboration is
invalid due to the model dependent prediction of LS of a
large E2 component in the CD of \b8, which
was ignored in our paper \cite{Mo93}.
In the following we show that their assertion arose from a
misunderstanding of the experimental procedures of the RIKEN experiment.

We first note that I have pointed to LS
that they have used and published
(before the RIKEN collaboration!) incorrect data, including wrong
error bar. Langanke and Shoppa now attempt
to correct it in a form of (a revised) Erratum.
Among several mistakes, LS appear to ignore the
angular resolution of the RIKEN experiment.
The finite angular resolution implies that the acceptance
(or response) of the RIKEN detector is not
the same for E1 and E2.  This invalidates the basic
assumption of LS that "Assumes the detector efficiency is the same
for E1 and E2 contributions" \cite{Lang1}. As it turns out the angular
averaging tends to push the predicted E1 cross section to large
angles, where the E2 dominates. And the large E2 predicted by LS
appears to be a compensation for their neglect of the angular
resolution of the RIKEN experiment. In that sense
the entire analysis of LS is misleading and in fact wrong.

A search for E2 component in the RIKEN data was performed by Gai
and Bertulani \cite{Ga94}. When the experimental
resolutions are correctly taken into account, together with the
correct RIKEN data (!) the best fit of the angular distributions is obtained
with E1 amplitude alone, as shown in Fig. 10. Our analysis invalidates
the claims of LS and support the analysis performed by Dr. Iwasa
and his advisor Professor Motobayashi, as presented in Ref. \cite{Mo93}.

In conclusion we demonstrate that the Coulomb dissociation provides
a viable alternative method for measuring small cross section of
interest for nuclear-astrophysics. First results on the CD of \b8 are
encouraging for a continued effort to extract $S_{17}(0)$, of importance
for the SSM. Our initial results are consistent with
the lower value of the cross section measured by Filippone et al. and
suggest a small value for the extracted $S_{17}(0)$; smaller than
20 eV-barn, and considerably smaller than assumed in the SSM of Bahcall
et al.

\section{Acknowledgements}

I would like to acknowledge the work of my graduate student Dr. Zhiping Zhao
who analyzed with great care the data on the beta-delayed alpha-particle
emission of $^{16}N$, a work for which she received the "1994 Dissertation
Award of the Division of Nuclear Physics of the American Physical Society".
In addition I acknowledge the work of
Dr. Naohita Iwasa and Professor
Tohru Motobayashi that performed the data analysis of the \b8
experiment at Rikkyo University; the Ph.D. thesis of Dr.
Iwasa.  I would also like to acknowledge discussions
and encouragements from Professor Carlos Bertulani,
Professor Gerhard Baur, and my collaborator Professor Thierry Delbar.

\newpage

\section{Figure Captions}

\underline{Fig. 1:}  Burning stages and onion-like structure of a $25M_\odot$
  star prior to its supernova explosion \cite{We80,Ro88}.\\
\   \\
\underline{Fig. 2:}  Nuclear States involved in the beta-delayed alpha-particle
   emission of $^{16}N$.\\
\   \\
\underline{Fig. 3:}  Clean Time of Flight Spectrum obtained in the Yale
   experiment.\\
\   \\
\underline{Fig. 4:}  (a) R-Matrix fit \cite{Ji90} of the spectrum measured at
   Yale \cite{Zh93,Zh93a,Zh93c} and (b) range of p-wave S-factors
   accepted by the measured data.\\
\   \\
\underline{Fig. 5:} Comparison of the spectra for beta-delayed alpha-particle
         emission of $^{16}N$ measured at Yale \cite{Zh93},
         TRIUMF \cite{Bu93} and Seattle \cite{Zh95}. \\
\   \\
\underline{Fig. 6:}  The $^{12}C(\alpha,\gamma)^{16}O$  cross section extracted
    from the observed solar abundances \cite{We93}.\\
\   \\
\underline{Fig. 7:}  The cross section of the Coulomb Dissociation as compared
    to the E1 capture cross section. \\
\   \\
\underline{Fig. 8:}  Radioactive beams extracted from the RIKEN-RIPS,
   used in the study of the Coulomb dissociation of $^8B$, a
   Rikkyo-RIKEN-Yale-Tokyo-LLN collaboration \cite{Mo93}.\\
\   \\
\underline{Fig. 9:} The measured S$_{17}$ factors for the \be7pg reaction,
   and extracted from Coulomb dissociation data \cite{Mo93}.\\
\   \\
\underline{Fig. 10:} The reduced $\chi ^2$ obtained from fitting the 600 keV
    angular distribution with: $\sigma_{CD}(E1)\ +\ \sigma_{CD}(E2)$.
    The best fit is obtained with E1 amplitude
    only, with $S_{17}=18\ eV-b$ (for the 600 keV
    angular distribution) \cite{Ga94}.

\end{document}